# Laser-assisted deposition of carbon nanotubes in optical fibers with multiparameter control

Ricardo E. da Silva and Cristiano M. B. Cordeiro

*Institute of Physics Gleb Wataghin, University of Campinas, Campinas, Brazil*

*Abstract*— We demonstrate a new method to deposit carbon nanotubes (CNT) on optical fibers based on a syringe-loaded CNT solution axially aligned to the fiber tip. A laser generates an optical tweezer in a water-based CNT solution, depositing nanotubes over the fiber cross-section. The parameters are adjusted, resulting in two deposited CNT layers with distinct thicknesses. This setup employs smaller solution volumes than those commonly used in beckers, providing high confinement, protection, and interaction of nanotubes, laser, and fiber, offering a promising alternative for real-time monitoring, which are significant to the development of industrial fiber lasers and biomedical optoacoustic devices.

*Keywords—laser-induced deposition, carbon nanotubes, optical fiber, optical tweezer.*

## I. INTRODUCTION

Carbon nanotubes (CNT) deposited on optical fibers enable remarkable applications as optical absorbers in passively mode-locked fiber lasers, all-optical switches, fiber sensors, and neurostimulators [1], [2], [3], [4]. CNTs offer high optical nonlinearity, diameter-dependent absorption bands, and chemical and mechanical stability. In fiber lasers, CNTs work as saturable absorbers, retaining and releasing laser power with an ultrafast recovery time (<1 ps), which is attractive for ultra-short pulse generation [5], [6]. Alternatively, CNT fiber-based devices offer potential features for ultrasonic neurostimulation to treat diseases [4]. A pulsed laser excites a CNT-polymer composite deposited at the fiber tip, which thermo-expands and generates highly confined ultrasonic waves [4]. Decreasing CNT layer thickness increases device frequency and bandwidth, improving spatial resolution in biomedical devices [4].

CNTs have been successfully deposited on fiber tips employing carbon wall papers poured into polystyrene cells after evaporation of a CNT solution over several days [7]. The 49.53 µm thick CNT paper is stripped out of the cell and attached to a fiber ferrule [7]. Alternatively, the dip coating methods require specific functionalization of nanotubes with organic compounds. The fiber tip is dipped in the CNT organic gel, and the deposited layer thickness is adjusted by changing the fiber dipping hold time, withdrawal velocity, and drying parameters. The deposited CNT layers are non-uniform with thicknesses less than 1 µm [8]. In contrast, laser-induced CNT deposition provides a faster method compared to previous techniques, allowing the use of simpler CNT solutions, based on water and alcohols [3]. In addition, this method offers a controllable spatial distribution of CNT agglomerates over the fiber cross-section [1], [2], [3].

In the laser-driven method, CNTs are usually deposited by vertically dipping an optical fiber in a CNT solution and injecting a high-power laser into the fiber [1], [2], [3], [6], [9]. The nanotubes are attracted by the laser beam and attached to the fiber tip by a combination of competitive physical mechanisms, such as thermophoresis (fluid thermal expansion), fluid convection, and radiation pressure, whose contributions depend on the fiber type, laser power magnitude and wavelength, CNT and solvent properties [9]. The resulting optical tweezer effect from laser beam traps nano- and micro-scaled particles by the spatial divergence of optical intensity at the cleaved fiber end [1], [10]. These potential non-uniform sources cause non-uniformly distributed CNT agglomerates on the fiber tip, with growing thickness by increasing laser power, deposition period, and CNT concentration [2], [3].

Previous studies have qualitatively investigated the thickness of deposited carbon layers by monitoring the reflected power at the fiber tip [2]. The increasing laser reflection over time is caused by the rising refractive index contrast with the accumulating CNTs being attached to the silica [2]. However, solution volume control, fiber alignment, and near-field monitoring of nanoparticle optical emissions have not yet been reported. In addition, the commonly employed centimeter-scale open beckers to store CNT solutions decrease the overlap of nanoparticles and laser beam, while exposing the solution to contaminants from the environment. Although fluid convection can be observed, optical emissions of CNT agglomerates may not be visible due to the large solution volumes surrounding the fiber. Besides, nanotube decantation, fiber misalignments, and tilts might reduce layer quality and efficiency during deposition [6].

We demonstrate a new method for depositing CNTs using an optically monitored, solution-controlled, and horizontally aligned syringe-fiber setup, which significantly enhances the interaction of CNT particles with the laser beam and the optical fiber. We fabricated CNT layers on SMFs with distinct distributions by using a water-based CNT solution. This study provides significant monitoring of the deposition process and an unprecedented view of CNT entanglements in the laser-induced optical tweezer.

## II. EXPERIMENTAL SETUP AND METHODS

Fig. 1(a) illustrates the experimental setup for depositing CNTs on the tip of a standard single-mode fiber (SMF). The CNT solution is based on an aqueous dispersion of 0.84 nm average diameter single-walled CNTs (≥ 99 %), resulting in a CNT concentration of about 0.12 mg/mL. Further details about the water-based CNT solution are found in [11], [12].

This work was supported by the grants 2022/10584-9, 2024/02995-4, São Paulo Research Foundation (FAPESP), 309989/2021-3, Conselho Nacional de Desenvolvimento Científico e Tecnológico (CNPq). We thank LIMicro-IQ Microscopy (RRID:SCR_024633), FAPESP grant (#2023/01620-4).

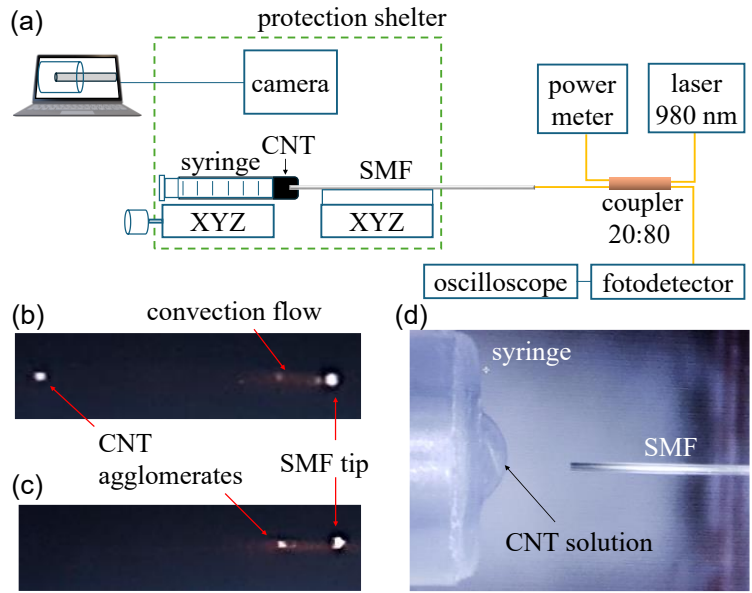

**Fig. 1.** (a) Illustration of the experimental setup used to deposit carbon nanotubes (CNT) in a standard single-mode optical fiber (SMF). The syringe and SMF are aligned with XYZ micro-positioners. (b)(c) Details of the SMF tip inside the syringe during deposition showing the laser attracting CNT agglomerates. (d) Details of the syringe and SMF tips after the CNT deposition.

We have deposited CNTs on SMFs by using two distinct solution volumes. The first sample, named SMF 1, employed a disposable syringe (with a maximum capacity of 1 mL) loaded with a 0.05 mL solution. The syringe is fixed on an XYZ micrometer positioner as indicated in Fig. 1(a). The fiber tip is cleaved and further cleaned by using low-power arc voltage with a fusion splicer. SMF is fixed on another XYZ positioner and aligned to the syringe tip with an objective camera connected to a laptop. The fiber is centrally aligned to the syringe's aperture and inserted into the CNT solution. A removable shelter covers the setup, blocking laser emissions through the syringe and contributing to isolating the components and the solution from contamination, illumination, and vibrations from the environment.

The power from the laser centered at 980 nm is split by the coupler, delivering more than 80 % at the SMF tip (about 88 mW - the laser wavelength and power levels are those employed in [3]). The remaining power reaching the power meter estimates the level at the fiber tip. The laser is turned ON for 5 minutes, and its reflection at the fiber tip is measured by a photodetector connected to an oscilloscope [2]. The camera is focused on the fiber tip, monitoring power emissions in the surrounding solution. Fig. 1(b) and 1(c) show emissions of arbitrary CNT agglomerates being attracted by the laser and deposited on the fiber. We noticed that the fiber end becomes brighter with increasing deposition time, indicating an increase in CNTs attached to the silica surface. Fig. 2(a) shows the laser reflection over deposition time. After the laser is ON, the reflection from the interface silica-CNT solution remains almost constant for about 3 minutes, increasing afterwards with CNT particles being attached to the fiber core. Fig. 1(d) shows a detail of the syringe and fiber tips after deposition. A water bubble is observed at the syringe tip, caused by the fluid thermal expansion (no solution has leaked out of the syringe even by applying higher powers).

The same setup and method are used for the second sample (SMF 2), employing a larger disposable syringe (3 mL maximum capacity) and CNT solution volume (2 mL) (the syringe needle tip was removed to reduce the bubble effect). The laser power is increased to about 139 mW, and deposition is monitored for 10 minutes. Fig. 2(b) shows the measured laser reflection, indicating increasing deposition after about 3 minutes. We note a reflection step probably caused by the strong rearrangement of CNT agglomerates over the fiber core. Power emissions of CNT clusters moving towards the fiber have also been observed.

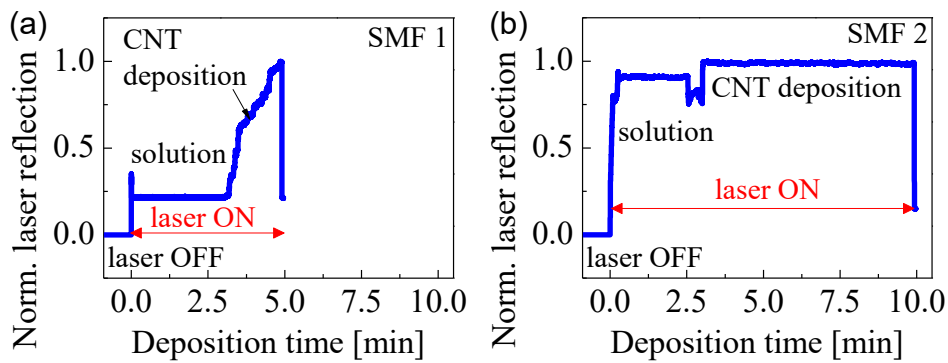

**Fig. 2.** (a) Laser reflection measured by the oscilloscope during the CNT deposition for the samples (a) SMF-CNT 1 and (b) SMF-CNT 2.

The effect of CNT depletion in the solution on the deposited layers is evaluated with two additional experiments. Two new samples (SMF 3 and 4) were respectively inserted into the CNT solutions previously depleted for SMF 1 and 2. The applied laser powers are respectively similar to those used for the previous samples, while the deposition periods are increased (SMF 3: 10 minutes, SMF 4: 17 minutes) to enhance deposition of the smaller amount of CNTs available in the solutions.

## III. RESULTS AND DISCUSSION

We have characterized the deposited CNT layers on the SMF samples by using a scanning electron microscope (SEM). The thickness and uniformity of the deposited layers could not be measured with our current available resources. Although the SEM microscope can measure localized deposited spots on the fiber side view (mainly on the fiber edges), the image spatial resolution depends on prior metal sputtering of samples (which can cover thin nanoparticle layers) and available microscope parameters and components. Thus, we could not detect thin CNT layers with nanoscale sizes mainly in the core region. In addition, SEM usually has limited scanning capability for evaluating deposition uniformity across the whole fiber cross-section. Contact profilometers with "needle"-like diameters compared to the fiber diameter are difficult to align on the fiber tip, having a maximum scanning step height (large non-uniformities on the deposited layer and the large step height at the fiber edges might damage the fiber tip and the profilometer probe). These issues become critical when using atomic force microscopes (AFM) with limited scanner $z$-axis travel and scan range. Overall, thickness measurements and analysis of CNT deposition uniformity on fiber tips are usually not reported, indicating these challenges might still be a broad limitation to compare quantitatively multiple fiber samples [1], [2], [3], [6].

Fig. 3(a) shows SMF 1 with a deposited layer that is apparently thinner and largely distributed over the fiber cross-section compared to previous studies using water-based CNT solutions [3]. It suggests that a volume-controlled CNT solution with high dispersion and low concentration of nanotubes might be suitable for improving uniform distribution. In addition, depositions are mostly distributed in the central region of the fiber cross-section compared to the fiber edges, indicating that the employed lower laser power might benefit thermophoresis and convection (attracting CNTs toward the laser beam axis), while reducing radiation pressure that deviates CNTs away from the fiber core [9]. Fig. 3(b) shows SMF 2 with a thicker layer deposited over the cladding, as shown by CNT agglomerates in the inset. It indicates that higher power, deposition time, and available CNT particles in the solution might contribute to increasing agglomerates compared to SMF 1. The higher CNT distribution on the cladding and edges rather than on the core indicates an increasing effect of radiation pressure, diverging particles from the laser beam axis (core) to the cladding [9]. These effects should also cause non-uniformities on SMF 3 and 4, respectively shown in Fig. 3(c) and 3(d). Reduction of deposited layers in the fiber core region should be caused by CNT depletion. The large dark spots observed in these samples can be formed by particles bundling in the solution or accumulating with longer deposition periods (CNTs on the fiber core region and cladding have been characterized using Raman spectroscopy, monitoring specific resonances of single-walled CNTs with a silica spectral background). Other sources of non-uniformities might come from distinct fiber types, laser beam profiles (favoring thermophoretic motion to hotter regions (beam axis) and convection outside the beam cone [9]), power fluctuations, and irregular thermal gradients induced by high density and inefficient thermal diffusion in the CNT-water interface [3].

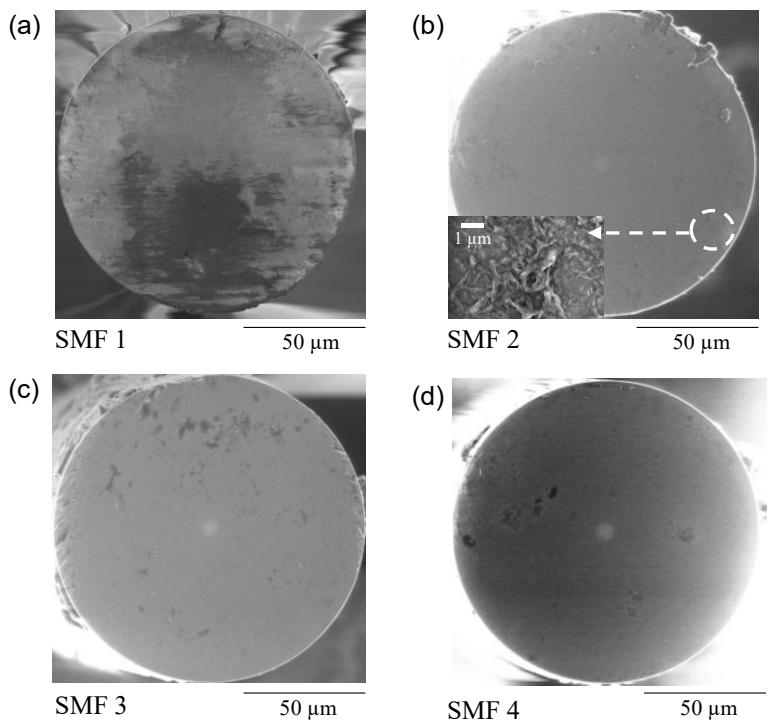

**Fig. 3**. Scanning electron microscope (SEM) images of the single-mode standard fibers (SMF) with deposited layers of carbon nanotubes: (a) SMF 1, (b) SMF 2 (inset: detail of CNT agglomerates), (c) SMF 3 and (d) SMF 4.

Overall, the results demonstrate the effect of solution volume control and CNT depletion. CNTs are more quickly depleted in the small volume (0.05 mL) than in the 2 mL solution, resulting in a more uniformly distributed layer as observed for SMF 1. This might contribute to decreasing non-uniform particle accumulation observed in water solutions using similar power and wavelength [3] (this improvement is also observed in the other SMF samples, mainly in the core region). It indicates that solution volume control is a useful parameter to limit undesirable effects caused by excessive power or prolonged deposition time. Further studies might consider water-based solutions with higher CNT concentrations. In addition, specific polymer-based CNT solutions might increase deposition thickness in the fiber core [1], [2], [3]. The use of transparent glass syringes should improve visibility of CNT emissions, enabling a useful tool to study optical tweezers. This technique might be applied to deposit CNTs also in microstructured fibers, photonic crystal fibers, and hollow-core fibers, which is promising for fiber sensors, fiber lasers, and optoacoustic ultrasonic transmitters.

## IV. CONCLUSION

We experimentally demonstrate the deposition of CNTs on optical fibers using a syringe and an axially aligned fiber for the first time. Four samples are fabricated by employing a water-based CNT solution and distinct setup parameters. Different layers are deposited on the SMF tip, employing SMF 1 with a small solution volume (0.05 mL), low power (88 mW), and 5 minutes deposition, and SMF 2 with a larger volume (2 mL), higher power (139 mW), and 10 minutes deposition. SMF 2 shows nanotube agglomerates, indicating a thicker deposited layer non-uniformly distributed on the fiber cladding. SMF 3 and 4 were prepared using the deployed CNT solutions and similar powers over longer times (10 and 17 minutes), showing that additional samples can be fabricated with small solution volumes, while reducing CNTs in the fiber core. Overall, this method and setup provided a high confinement of CNT solution around the fiber tip, significantly increasing the interaction of nanoparticles and the laser. This is promising to monitor and study laser-induced optical tweezers, increasing deposition efficiency while reducing material and cost.

## ACKNOWLEDGMENT

We thank Prof. João Paulo Vita Damasceno with the Group of Nano Solids (GNS) at UNICAMP for kindly providing the CNT solution. We thank the groups with LAMULT (IFGW) and LIMicro (IQ), especially Rosane Palissari, Bruno Camarero Guidi, and Ana Letícia Moreira da Fonseca, for the support with the Raman spectroscopy and SEM images.